\documentclass[12pt]{elsarticle}
\usepackage[margin=.75in]{geometry}
\usepackage{multirow}
\usepackage{graphicx}
\usepackage{caption}
\usepackage{booktabs}
\usepackage{subcaption}
\usepackage{mathtools}
\usepackage{textcomp}
\usepackage{floatrow}
\usepackage{svg}
\DeclareCaptionLabelFormat{andtable}{#1~#2  \&  \tablename~\thetable}
\usepackage{lineno}

\usepackage{natbib}
\usepackage{lineno,hyperref}
\usepackage{color,soul}
\usepackage{floatrow}
\usepackage{multicol}
\usepackage{lscape}
\usepackage{amsmath,amssymb}
\makeatletter
\def\ps@pprintTitle{%
   \let\@oddhead\@empty
   \let\@evenhead\@empty
   \let\@oddfoot\@empty
   \let\@evenfoot\@oddfoot
}
\makeatother
\newfloatcommand{capbtabbox}{table}[][\FBwidth]
\usepackage[ruled,vlined]{algorithm2e}

\biboptions{sort&compress}

\begin{document}

\begin{frontmatter}

\title{Generalized Reinforcement Learning for Building Control using Behavioral Cloning}

\author{Zachary E. Lee}
\author{K. Max Zhang}

\address{Sibley School of Mechanical and Aerospace Engineering, Cornell University, Ithaca, NY 14853, USA}
\cortext[correspondent]{Corresponding author: kz33@cornell.edu}

\begin{abstract}

Advanced building control methods such as model predictive control (MPC) offer significant potential benefits to both consumers and grid operators, but the high computational requirements have acted as barriers to more widespread adoption. Local control computation requires installation of expensive computational hardware, while cloud computing introduces data security and privacy concerns. In this paper, we drastically reduce the local computational requirements of advanced building control through a reinforcement learning (RL)-based approach called Behavioral Cloning, which represents the MPC policy as a neural network that can be locally implemented and quickly computed on a low-cost programmable logic controller. While previous RL and approximate MPC methods must be specifically trained for each building, our key improvement is that our controller can generalize to many buildings, electricity rates, and thermostat setpoint schedules without additional, effort-intensive retraining. To provide this versatility, we have adapted the traditional Behavioral Cloning approach through (1) a constraint-informed parameter grouping (CIPG) method that provides a more efficient representation of the training data; (2) an MPC-Guided training data generation method using the DAgger algorithm that improves stability and constraint satisfaction; and (3) a new deep learning model-structure called reverse-time recurrent neural networks (RT-RNN) that allows future information to flow backward in time to more effectively interpret the temporal information in disturbance predictions. The result is an easy-to-deploy, generalized behavioral clone of MPC that can be implemented on a programmable logic controller and requires little building-specific controller tuning, reducing the effort and costs associated with implementing smart residential heat pump control.

\end{abstract}


\end{frontmatter}
\section*{Highlights}

\begin{itemize}
\item Behavioral Cloning reduces model predictive control (MPC) computational requirements.
\item One Behavioral Cloning agent controls different buildings without further training.
\item Constraint-informed parameter groupings provide more efficient state representations.
\item Reverse-Time Recurrent Neural Networks incorporate future disturbance predictions.
\item Simulations show Behavioral Cloning offers energy efficient control similar to MPC.
\end{itemize}

\noindent\makebox[\linewidth]{\rule{\textwidth}{0.4pt}}

\section{Introduction}
\linespread{1.25}\selectfont

As the energy system relies more and more on variable renewable energy sources, efficient grid-interactive buildings that can modulate their demand according to the availability of renewable energy become ever more important. Buildings are becoming increasingly electrified, replacing fossil fuel based space heating with clean, electric alternatives such as heat pumps. A substantial amount of research has shown that smart heat pump control can harness the inherent thermal storage of the building envelope and provide important grid services such as load shifting and demand response \cite{Lee2020}, which are generally considered as requirements for maintaining a reliable electrical grid with high penetrations of renewable energy resources \cite{Shaner2018}. However, more advanced control methods that can provide this demand flexibility still have large technical, economical, and social barriers to adoption, and therefore they lack the scalability needed to have a large impact on the overall energy system.

One of the most widely studied advanced building control methods is model predictive control (MPC) \cite{AFRAM2016}. Compared to conventional rule-based approaches, MPC can offer substantial energy consumption savings of 20\% or more \cite{serale2018model}, as well as achieve other control objectives such as peak load reduction \cite{Lee2020a} and demand response \cite{finck2019economic}. At each time step, a constrained optimization problem is solved to determine the optimal control given a model of the building and predictions of future disturbances like weather, occupancy, and electricity prices. But despite substantial research efforts into the development of MPC for heat pumps, it has yet to be widely adopted due to its costly installation and computational hardware costs \cite{cigler2013beyond}. With over 30\% of US households already reporting some difficulty in paying their energy bills \cite{RECS}, these high capital costs can make advanced building control economically unfeasible for low-income populations and neglect a significant source of demand flexibility.

One potential avenue for more scalable building control is through smart thermostats, which feature a simple plug-and-play installation that has resulted in rapid recent adoption \cite{berg2017}. However, smart thermostats have limited computational hardware that often cannot handle the high memory and processing requirements needed to solve MPC. Instead, smart thermostats use rule-based approaches for energy efficiency and demand response, but can connect to the cloud for more advanced data processing. While they can reduce energy consumption, these rule-based control methods often provide insufficient perceived benefit to justify the high capital costs of smart thermostats. In a recent US nationwide survey, 30\% of people said that smart thermostats are too expensive and 60\% said that they simply do not see the merits of upgrading their current system \cite{deloitte}. Moreover, for cloud-based smart thermostats, data privacy and security are other key concerns \cite{hernandez2014smart}. Thus, inexpensive plug-and-play control solutions that provide higher cost savings while preserving data privacy can reduce many of the barriers to more widespread adoption of smart heat pump control.

Some studies have proposed simplifying the online MPC computation so that it can be computed locally on low-cost, resource-constrained devices such as programmable logic controllers (PLC), which typically have limited processing speeds in the range of MHz and memory on the order of hundreds of kB. One method of simplifying MPC is through deriving an explicit representation of the optimal MPC policy. Explicit stochastic MPC \cite{Drgona2013} and explicit scenario-based MPC \cite{Parisio2014} have both been shown to drastically improve MPC computational time for building climate control for small state spaces and time horizons. However, as the problem size gets larger, explicit MPC approaches require substantially more memory, meaning that the longer time horizons needed to achieve the benefits of building MPC can be prohibitively large. A memory-efficient approach called approximate MPC (AMPC) was recently proposed by Drgona\v{n}a et al. \cite{drgovna2018approximate} that approximated the explicit MPC formulation by training a feed-forward neural network on samples generated from closed loop MPC simulations for a single building. However, by being specific to a single building, its implementation would require building model development and new MPC simulations to train each new building-specific AMPC model. This requirement can amplify the effort needed for model and controller development, an already challenging problem for implementing conventional MPC.

More recently, reinforcement learning (RL) has emerged as an extremely powerful tool for learning optimal control policies for systems with large state spaces. Rather than analytically deriving the mapping from the state to the optimal control, RL leverages machine learning techniques like deep learning to learn an approximate mapping by interacting with the system. Deep Q-learning was used in \cite{Wei2017rl} to control a commercial building by learning through interaction with a model built in EnergyPlus \cite{Crawley00energyplus:energy}. Similarly, in \cite{ZHANG2019472}, the use of building energy modeling software like EnergyPlus for training RL controllers was further explored using the actor critic method \cite{mnih2016asynchronous}. Like most other RL approaches, these studies can require significant training time through interaction with either a simulated building, which can be expensive to develop, or the real building, which can be expensive to experiment on.

While both AMPC and RL-based prior approaches are able to learn complex optimal control policies for large problems and require very little online computational hardware, they both face the same barrier to more widespread implementation: They are specifically trained for a single building and must be retrained for each additional installation, introducing significantly more installation costs. Thus by being unable to generalize, these approaches have limited scalability, particularly for smaller buildings with less ability to cover capital costs \cite{WANG2020115036}.

In this paper, we combine ideas from both AMPC and RL research to create a generalized MPC-like controller that can control many buildings using resource-constrained devices while being trained only once. In particular, we apply a form of RL called \textit{Behavioral Cloning} \cite{attia2018global} that has to our knowledge not been applied in the building control context. Behavioral Cloning attempts to mimic the actions of an available expert controller, such as MPC, and results in much faster training compared to other RL approaches \cite{zhang2016queryefficient}. As a result, we can train the controller on a wide range of buildings, setpoint schedules, and electricity rates and so that it can generalize to different buildings, various resident preferences, and changing utility prices without additional controller tuning. 

In addition, we improve the conventional Behavioral Cloning approach to make it more suitable for building control. First, we introduce a more efficient representation of the input state using MPC constraint-informed parameter groupings (CIPG). Second, we present a new machine learning model structure called reverse-time recurrent neural networks (RT-RNN) that more accurately interprets the future disturbance information that is vital to effectively pre-heat or pre-cool the building for energy efficiency. The result is a behavioral clone of MPC that uses only around 100 kB of memory, requires negligible computational time, and can be implemented in buildings on a PLC in a low-cost thermostat with minimal installation effort and costs.

The paper is organized as follows. Section \ref{sec: methodology} gives an overview of the paper's methodology. Section \ref{sec: mpc} provides the system model and MPC formulation that serves as the expert controller. Section \ref{sec: appr} describes how Behavioral Cloning is developed and evaluated, including the constraint-specific parameter normalization and DAgger training data generation algorithm. Section \ref{sec: results} presents the models' control results and computational requirements. Section \ref{sec: conclusion} concludes the paper.

\section{Methodology}
\label{sec: methodology}

\begin{figure}[b]
    \centering
    \includegraphics[width=.85\textwidth]{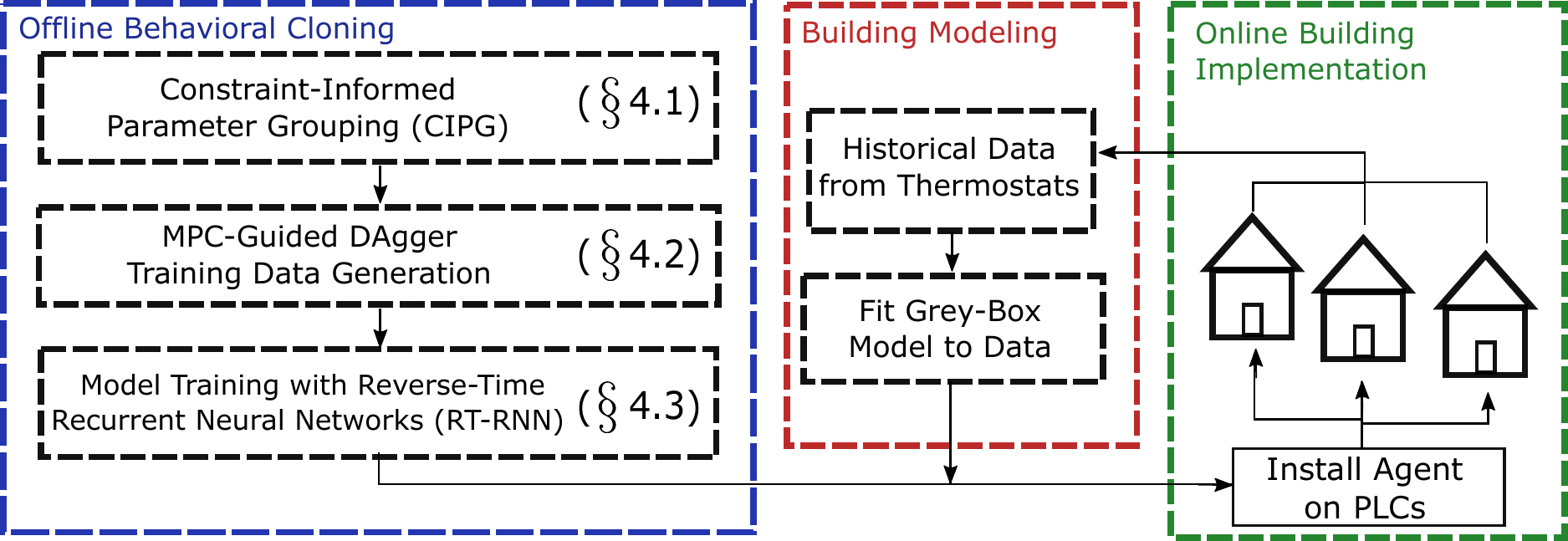}
    \caption{Overall methodology for Behavioral Cloning of MPC and implementing it in a population of buildings. Behavioral Cloning and building modeling can both be performed offline using a desktop computer, while online control only requires a programmable logic controller (PLC).}
    \label{fig: methodology}
\end{figure}

Our new methodology for Behavioral Cloning of building MPC is outlined in Fig. \ref{fig: methodology}. Since we require the Behavioral Cloning agent to generalize to new buildings and operating conditions, the training process is significantly more difficult than if the agent were trained for a specific building. Therefore, we present three main contributions that improve the agent's ability to generalize when compared to the AMPC approach.

First, we developed a method called \textit{constraint-informed parameter groupings (CIPG)} to provide a more efficient representation of the state and disturbance inputs. These parameter groupings are derived from the structure of the MPC constraints and building model. Rather than using a black-box or white-box model, which can sometimes require hundreds of unique parameters, we use a reduced-order grey-box building model to allow the thermodynamics to be efficiently grouped as inputs to the Behavioral Cloning agent. By exploiting the structure of the MPC formulation, we group the model and disturbance parameters in a way that provides a more efficient representation of the input state information and improves the ability for the controller to generalize to new conditions. In essence, the building thermodynamics and the optimal control are not necessarily functions of the actual parameter values, but rather the ratios or differences between the parameter values (e.g., heat loss is a function of the temperature difference). Therefore, by grouping the training data parameters based on the constraints in the MPC formulation, we condense the feature space to allow operating conditions from one training simulation to be more effectively applied to a different operating condition during test time. More description of CIPG can be found in Sec. \ref{sec: CIPG}.

Second, we build a diverse set of training data that provides the Behavioral Cloning agent with sufficient information to operate optimally under various operating conditions and to recover from any mistakes. Since the agent is sub-optimal, it will likely make errors in its control prediction, which can lead it to a state that differs from the expert controller's trajectory. Since these sub-optimal states would not be included in training data generated by the expert controller, the agent would have no prior information on which to base its control decision and will likely make more errors, leading to instability and constraint violation. To prevent compounding errors, we use the \textit{Dataset Aggregation algorithm (DAgger)} \cite{ross2011reduction} guided by MPC as the expert controller. At each iteration, DAgger uses the Behavioral Cloning agent to control a simulation of randomized building models and operating conditions over a period of time. At each of these time steps, the MPC solves and records the true optimal control and adds the new state-control pairs to the training dataset. After each iteration, the agent is retrained with the growing dataset that now has information on how to correct the mistakes the agent made during the prior iteration. This process is repeated until each iteration produces negligible performance improvements. This process is described in Sec. \ref{sec: DAgger}.

Third, the total grouped and normalized dataset is used to train the final Behavioral Cloning agent that predicts the optimal control as a function of the input parameters. We propose a new model structure elaborated in Sec. \ref{sec: RT-RNN} called \textit{reverse-time recurrent neural networks (RT-RNN)} that can more effectively incorporate the temporal information from the future disturbance forecasts compared to more conventional supervised learning algorithms. We test our model structure against three other supervised learning methods to show that RT-RNNs can provide the best performance while maintaining minimal memory and processing requirements.

Finally, the optimal model is implemented in a test simulation on a sample of buildings intended to mimic real-world operation. To implement the controller, a homeowner buys and installs a low-cost thermostat containing a PLC with the Behavioral Cloning agent installed. The thermostat then collects various operational data over a period of time that can be used to automatically derive a data-driven reduced order building model using the method given in \cite{Lee2020a}. These model parameters, combined with weather forecasts and data collected by the thermostat, are then used as inputs to the Behavioral Cloning agent to provide online approximately optimal control. This test simulation contains buildings with diverse thermodynamics and heat pump performances, various thermostat setpoint schedules obtained from real data, and different electricity price schedules, all of which were not originally included in the training dataset. By testing on these diverse operating conditions, we show our Behavioral Cloning approach leads to improved versatility and minimal-effort implementation compared to the current state-of-the-art building AMPC \cite{drgovna2018approximate}.

\section{Model Predictive Control}
\label{sec: mpc}

\subsection{Model Definition}

Our methodology is designed to be applicable to a wide range of residential building types and heat pump configurations. In this paper, we simulate one of the most common residential configurations: a detached home served by a single-stage air-to-air heat pump. However, many heat pump MPC formulations consist of similar structures, and thus it is straightforward to adapt our methodology to other system types for both heating and cooling.

To be able to model each system without significant manual effort, we use a data-driven grey-box model, where each of the building and heat pump model parameters can be automatically identified from collected data. As a reduced-order model, the grey-box approach reduces the number of thermodynamic model parameters so they can be used as inputs to the Behavioral Cloning agent. While deriving these building parameters sometimes require data collected from a variety of sensors throughout the building \cite{date2016} or from building energy simulations \cite{Eisenhower2012}, we use the identification and control method presented in \cite{Lee2020a}, which is designed to require minimal hardware installation cost and effort.

The building model can be represented by a thermal resistance-capacitance (RC) circuit. While single-state RC models often have insufficient complexity to provide benefits from MPC \cite{Blum2019}, a two-state model better captures the increased energy storage capacity of the building's construction and is applicable to a wide range of buildings. This two-state model includes different states for the building's indoor air and the building's construction and includes effects from solar irradiation, given by \cite{Lee2020a},

\begin{equation}
\begin{aligned}
& C_a \Dot{T}_a(t) = \frac{T_\infty - T_a(t)}{R_{a\infty}} + \frac{T_m(t) - T_a(t)}{R_{am}} + \alpha_a G  + Q_\text{HP} \\
& C_m \Dot{T}_m(t) = \frac{T_\infty - T_m(t)}{R_{m\infty}} + \frac{T_a(t) - T_m(t)}{R_{am}} + \alpha_m G,
\label{3R2C}
\end{aligned}
\end{equation}

\noindent where the subscript $a$ refers to the indoor air, $m$  to the building mass, and $\infty$ to the outside air. The resistance and capacitance values are given by $R$ and $C$, respectively, while the temperature of the states is given by $T$. Solar heat gains are included using the solar irradiation $G$ and the solar absorption factor $\alpha$. Based on an analysis of manufacturer performance data \cite{TH4B}, the heat transfer from the heat pump $Q_\text{HP}$ are assumed to vary linearly based on the indoor and outdoor temperature, given by,

\begin{equation}
Q_{\text{HP},a} = u(\beta_1 (T_{\infty} - T_{a}) + \beta_2).
\label{LinearModels}
\end{equation}

\noindent where $\beta_i$ are data-driven heat pump specific model parameters and $u$ denotes the binary control input for whether the heat pump is on or off.

Finally, based on an analysis of data in \cite{TH4B} the power consumption $P$ is assumed to be a constant $\gamma$ multiplied by the control input,

\begin{equation*}
\begin{aligned}
 P = \gamma u
  \end{aligned}
\end{equation*}

\noindent For use in MPC, the model is discretized with time step $\Delta t$ into the state space form indexed by $k$,

\begin{equation}
    \boldsymbol{x}_{k+1} = A \boldsymbol{x}_k + B_k \boldsymbol{u}_k + E \boldsymbol{w}_k,
    \label{eq: state_space}
\end{equation}

\noindent where,

\begin{equation*}
\begin{aligned}
    & \boldsymbol{x} = \Big[\begin{matrix}T_{a,k} \\ T_{m,k}\end{matrix}\Big], \quad 
    \boldsymbol{u_k} = \Big[\begin{matrix}u_k\end{matrix}\Big], \quad
    \boldsymbol{w_k} = \Bigg[\begin{matrix}T_{\infty,k} \\ G_k\end{matrix}\Bigg] \\
    & A = \Bigg[\begin{matrix}1 - \frac{\Delta t}{C_a} \Big(\frac{1}{R_{a\infty}} + \frac{1}{R_{am}}\Big) & \frac{\Delta t}{C_a R_{am}} \\ \frac{\Delta t}{C_m R_{am}} & 1 - \frac{\Delta t}{C_m} \Big(\frac{1}{R_{m\infty}} + \frac{1}{R_{am}}\Big) & \end{matrix}\Bigg], \\
    & B_k =  \Bigg[\begin{matrix}\frac{\Delta t}{C_a}\Big(\beta_1(T_{\infty,k} - T_{\text{set}, k}) + \beta_2\Big) \\ 0 \end{matrix}\Bigg], \\
    & E =  \Bigg[\begin{matrix}\frac{\Delta t}{R_{a\infty} C_a}  & \frac{\alpha_a \Delta t}{C_a} \vspace{1mm}\\ \frac{\Delta t}{R_{m\infty} C_m}  & \frac{\alpha_m \Delta t}{C_m} \end{matrix}\Bigg].  
\end{aligned}
\end{equation*}

\subsection{Control Formulation}

The controller seeks to minimize the time-varying electricity cost while maintaining thermal comfort in response to varying thermostat setpoints decided by the resident. We define thermal comfort as a temperature range above and below the thermostat setpoint. Since we assume that setpoints are customizable by the resident, to maintain feasibility we penalize violations outside of this thermal comfort band. These violations are enforced by the constraints, 
     
\begin{equation}
\begin{aligned}
     T_{k+j} \leq T_{\text{set},k+j} + T_{\delta,k+j} + \overline{T}_{\text{pen},k+j}  \quad \forall j \in N\\
 T_{k+j}  \geq T_{\text{set},k+j} - T_{\delta,k+j}^i - \underline{T}_{\text{pen},k+j}   \quad \forall j \in N.  
 \end{aligned}
\end{equation}

Here, $ T_{pen,k+j} $ is the comfort violation decision variable,  $ T_{\delta,k+j} $ is the resident’s specified comfort band above or below the setpoint, $T_{\text{set},k+j}$ is the resident's specified setpoint, and $N$ represents the prediction horizon indexed by $j$. Note that the comfort band can also vary based on time of day and can be determined by whether the thermostat is in home, away, or sleep modes.

Next, heat pumps have inherent minimum on and off times to prevent short cyling and the resulting compressor damage and efficiency reduction. To enforce these minimum cycle times, we add the following constraints,

\begin{equation}
\begin{aligned}
     u_{k+j} - u_{k+j-1} = v_{k+j}^{\uparrow} - v_{k+j}^{\downarrow} \quad \forall j \in N
 \end{aligned}
\end{equation}
\begin{equation}
\begin{aligned}
     \sum_{i=k+j-t_{\text{min on}}}^{k+j} v_i^{\uparrow} \leq u_{k+j} \quad \forall j \in N
 \end{aligned}
\end{equation}
\begin{equation}
\begin{aligned}
     \sum_{i=k+j-t_{\text{min off}}}^{k+j} v_i^{\downarrow} \leq 1-u_{k+j} \quad \forall j \in N.
 \end{aligned}
\end{equation}

\noindent Here, $v_i^{\uparrow}$ and $ v_i^{\downarrow}$ are binary variables that are unity if the heat pump turned on or off, respectively, at the time step $i$. The parameters $t_\text{min, on}$ and $t_\text{min, off}$ are the minimum on and off times, respectively.

The objective function combines the time-varying cost of electricity $\pi_{e.j}$ with the upper and lower thermal comfort penalties, $\underline{\pi}_{\text{pen}}$ and $\overline{\pi}_{\text{pen}}$,

\begin{equation}
\begin{aligned}
 \underset{u_{k+j}}{\text{min}} \quad J = \sum_{j=0}^{N-1} \bigl[\pi_{e,k+j} P_{k+j} + \underline{\pi}_{\text{pen}}\underline{T}_{\text{pen},k+j} + \overline{\pi}_{\text{pen}}\overline{T}_{\text{pen},k+j}\bigr]
  \end{aligned}
\end{equation}

\noindent The final MPC problem is therefore,
 
 \begin{subequations}
\begin{align}
    & \quad \underset{u_{k+j}}{\text{min}} \quad \sum_{j=0}^{N-1} \bigl[\pi_{e,k+j} P_{k+j} + \underline{\pi}_{\text{pen}}\underline{T}_{\text{pen},k+j} + \overline{\pi}_{\text{pen}}\overline{T}_{\text{pen},k+j}\bigr] \\
    \nonumber \text{subject to } \\
    & \quad \boldsymbol{x}_{k+j+1} = A \boldsymbol{x}_{k+j} + B_{k+j} \boldsymbol{u}_{k+j} + E \boldsymbol{w}_{k+j} \quad \forall j \in N 
        \label{eq:ss} \\
    & \quad T_{a,k+j} \leq T_{\text{set},k+j} + T_{\delta,k+j} + \overline{T}_{\text{pen},k+j}  \quad \forall j \in N
        \\
    & \quad T_{a,k+j}  \geq T_{\text{set},k+j} - T_{\delta,k+j}^i - \underline{T}_{\text{pen},k+j}   \quad \forall j \in N        
        \label{eq: bounds}\\
    & \quad      u_{k+j} - u_{k+j-1} = v_{k+j}^{\uparrow} - v_{k+j}^{\downarrow} \quad \forall j \in N 
        \label{eq: min off 1} \\
    & \quad           \sum_{i=k+j-t_{\text{min on}}}^{k+j} v_i^{\uparrow} \leq u_{k+j} \quad \forall j \in N 
        \label{eq: min off 2} \\
    & \quad      \sum_{i=k+j-t_{\text{min off}}}^{k+j} v_i^{\downarrow} \leq 1-u_{k+j} \quad \forall j \in N. 
        \label{eq: min off 3} \\
\end{align}
\label{eq: complete model}
\end{subequations}
 
This gives the optimal MPC policy $\mu^*_{\text{mpc}}$ that maps the building parameters and disturbance forecasts to the optimal control $u^*_k$,

$$ u^*_k = \mu^*_{\text{MPC}}(\boldsymbol{x}_k, A, B_{k+j}, E, \gamma, T_{\text{set},k+j}, T_{\delta,k+j}, \pi_{e,k+j}, \underline{\pi}_{\text{pen}}, \overline{\pi}_{\text{pen}}). $$

\noindent where $\mu^*(\cdot)$ is found numerically by solving the optimization problem.

\section{Learning the MPC policy}
\label{sec: appr}

Solving this mixed-integer linear program can require significant processing power and memory. Therefore, we seek to use Behavioral Cloning to find the function $\hat{\mu}$ that represents the MPC policy such that it provides the optimal control at each time step as a function of the inputs to the MPC. Thus, Behavioral Cloning takes the form,

$$ u_k = \hat{\mu}_{\text{MPC}} (\cdot). $$

Finding this function is a supervised learning problem that uses the MPC controller to generate a diverse set of input parameters and the corresponding optimal control. Required input features to the Behavioral Cloning agent should include the same information used by the MPC, including future thermostat setpoints, comfort levels, electricity prices, and weather forecasts, as well as the building and heat pump model parameters. The agent's output is the binary optimal control value at the current timestep $u_k$ indicating whether the heat pump should turn on or off.

\subsection{Constraint Informed Parameter Groupings (CIPG)}
\label{sec: CIPG}
Since the controller will be implemented on a variety of home types, setpoint schedules, electricity tariffs, and weather forecasts, learning over the entire feasible parameter space can require a large amount of training data. Therefore, we propose constraint-informed parameter groupings (CIPG). CIPG groups the input parameters using knowledge of the MPC constraints to create more sample-efficient state representation in training data and improve controller performance on new conditions.

Our method to group the features is inspired by dimensionality reduction ideas from of the Buckingham Pi Theorem. As an example, this method is used in fluid dynamics to non-dimensionalize fluid parameters such that the solutions to complex fluid flows are no longer functions of the actual parameter values (e.g., viscosity, velocity, temperature, etc.), but instead functions of the ratios between the values (e.g., Reynolds number). This method can be similarly applied to normalize our input parameters using knowledge of the structure of the dynamics and MPC formulation. For example, heat loss is not an explicit function of the outdoor temperature, but rather the difference between the indoor and outdoor temperatures and the building's thermal parameters. Similarly, the control decision does not change if the costs were all scaled by 50\%. Therefore, grouping and normalizing these values provides a more efficient state representation without sacrificing any information required to solve the control problem. Through these groupings the training data for one building and operating condition can generalize to another, as long as these groupings remain the same.

The specific procedure for parameter grouping is described as follows. First, the building's thermodynamic parameters $R$, $C$ are simply grouped as the entries of the $A$ state space matrix defined in Eq. \ref{eq: state_space}. Though this initial grouping is quite straightforward, it illustrates the point that it is not the parameter values themselves that govern the MPC solution, but the ratios of the parameters instead. Following the notation of the Buckingham Pi Theorem where $\Pi$ refers to a grouped parameter, the building model parameter groupings are given by the vector,

\begin{equation}
    \Pi_1 = [a_{11}, a_{12}, a_{21}, a_{22}],
\end{equation}

\noindent where the subscripts denote the entries in the corresponding state space matrix.

Next, the heat pump's effect on the indoor air temperature comes from the $B$ matrix defined in Eq. \ref{eq: state_space}. Since the heat output changes based on the indoor and outdoor air temperature, this parameter grouping is indexed by $j$ over the MPC horizon $N$. The normalized parameter corresponding to the heat pump is given as,

\begin{equation}
    \Pi_{2,k+j} = b_{11,k+j} \quad \forall j \in N.
\end{equation}

The weather's effect on the solution comes from the forecasts for outdoor temperature and solar irradiation and the corresponding thermal properties of the home grouped in the $C$ matrix defined in Eq. \ref{eq: state_space}. We combine these into a matrix indexed over the MPC horizon,

\begin{equation}
    \Pi_{3,k+j} = \begin{bmatrix}c_{11} T_{\infty,k+j} \\ c_{21} T_{\infty,k+j} \\ c_{12} G_{k+j} \\ c_{22} G_{k+j}\end{bmatrix}, \quad \forall j \in N.
\end{equation}

We normalize the thermal comfort constraints by taking the distance between the temperature at the current time step, $T_{a,0}$, and the upper and lower thermal comfort bounds indexed over the control horizon. This grouping is defined such that value will be zero if $T_{a,0}$ is at the lower comfort bound and unity if it is at the upper comfort bound, given by,

\begin{equation}
    \Pi_{4,k+j} = \frac{T_{a,k} - (T_{\text{set},k+j} - T_{\delta,k+j})}{2 T_{\delta,k+j}} \quad \forall j \in N.
\end{equation}

Similarly, the normalized parameter corresponding to electricity price is the distance between the electricity price at the current time step and the maximum and minimum electricity prices such that the value is zero at the minimum price, and one at the maximum price. It is then multiplied by $\gamma$ to give the total energy cost of turning the heat pump on. This grouping is indexed over the MPC horizon and given by,

\begin{equation}
    \Pi_{5,k+j} = \gamma \frac{\pi_{e,k+j} - \pi_{e,\text{min}}}{\pi_{e,\text{max}} - \pi_{e,\text{min}}} \quad \forall j \in N.
\end{equation}

Finally, we implement the minimum heat pump on and off time constraints by supplying the previous control values. Since we assume a 15-minute minimum heat pump cycle time and a five minute time step, this becomes three previous control steps,

\begin{equation}
    \Pi_{6, k} = [u_{k-1}, u_{k-2}, u_{k-3}]
\end{equation}

The result is a new functional form for the approximate MPC policy that is a function of the normalized parameter groupings and spans a reduced parameter space,

\begin{equation}
   u^* = \hat{\mu}_\text{MPC}(\Pi_1,  \Pi_{2,k}, \Pi_{3,k}, \Pi_{4,k}, \Pi_{5,k},  \Pi_{6, k}).
\end{equation}

\subsection{Model Training using MPC-Guided DAgger}
\label{sec: DAgger}
Since the original controller is completely replaced by a machine learning model, the original MPC properties of recursive feasibility and closed loop stability are no longer guaranteed. Instead, the controller must be able to learn these properties solely from the training data, making the training data's content immensely important for stability and performance. For some applications such as in \cite{drgovna2018approximate} where the operating conditions do not vary much, the control prediction can make very few errors and can provide statistical guarantees on constraint satisfaction and stability \cite{Hertneck2018}. However, if the agent faces new operating conditions, such as changing setpoints or different electricity tariffs, the agent will likely make mistakes and deviate from the optimal control trajectory. Therefore, closed loop MPC simulations are often unable to provide sufficient information for the agent to correct itself should it drift to a sub-optimal state (such as outside of the thermal comfort bounds).

One of the most common algorithms for generating additional data outside of the optimal control trajectory is called Dataset Aggregation (DAgger) \cite{ross2011reduction}. DAgger is used in many different Behavioral Cloning applications, ranging from natural language processing \cite{vlachos2013investigation} to autonomous driving \cite{zhang2016queryefficient}. DAgger is an iterative algorithm that uses the Behavioral Cloning agent to generate likely suboptimal trajectories, records what the expert controller would have done in those trajectories, and then retrains the agent with additional data. We use MPC to guide the DAgger algorithm, which lets the suboptimal agent control the system, while the true MPC acts as a guide to correct the suboptimal behavior between each iteration. For example, if the agent lets the temperature drift below the lower thermal comfort bound, the true MPC will record what the optimal control should have been and how to correct for it so that the mistake is not repeated during future iterations. By letting the true MPC guide the imperfect Behavioral Cloning agent, DAgger enriches the training dataset above pure closed loop MPC simulations to allow the agent to be stable on new operating conditions and correct for model imperfection.

Our implementation of the DAgger algorithm starts by training an initial Behavioral Cloning agent on a day of closed loop MPC simulation data on a set of random building and heat pump model parameters. This initial agent is then tested in simulation to control a new set of random buildings on a new day with different weather conditions and electricity costs. During this initial test simulation, the agent will likely perform poorly and deviate from the optimal control trajectory. Throughout the test, however, a supervisory MPC calculates and records, but does not implement, the true optimal control at each time step. At the end of the simulation, these optimal control solutions are added to the training data set and the agent is retrained with the additional data. Through this process, the correct control responses to suboptimal states are added to the training dataset so the agent can know how to correct itself in the future. These iterations can be repeated until the agent is stable during the testing phase and its objective value $J_\text{appr}$ is within some limit $\epsilon$ of the true MPC objective value $ J_\text{MPC}$. The algorithm is given below,
\vspace{4mm}
\begin{algorithm}[H]
\SetAlgoLined
 Simulate a set of randomized buildings, weather, and electricity tariffs using MPC\;
 Train Behavioral Cloning agent on resulting normalized dataset\;
 \While{$ J_\text{appr} - J_{\text{MPC}} \leq \varepsilon $}{
  Simulate new set of randomized buildings, weather, and electricity tariffs controlled using the agent\;
  At each time step, solve MPC and add (but do not implement) the inputs and solutions to training dataset\;
  Retrain agent with additional training data\;
  Evaluate agent on test conditions and calculate total objective value $J_\text{appr}$\;
  }
 \caption{MPC-Guided DAgger}
\end{algorithm}

\bigskip

\subsection{Behavioral cloning model structure}
\label{sec: RT-RNN}
The type of machine learning model structure is especially important to develop a functioning Behavioral Cloning agent. Particularly with MPC, the ability to extract the temporal information embedded in the disturbance forecasts heavily affects the model's performance. For example, knowing that the setpoint will rise at a specific time in the future determines at what time the agent should begin preheating. Conventional supervised learning techniques previously used in approximate MPC \cite{drgovna2018approximate} like regression trees and feed-forward neural networks do not contain any inherent structure to interpret temporal information and thus were not sufficient to learn the larger feature space and be able to generalize to new conditions. Therefore, we propose a new model structure called reverse-time recurrent neural networks (RT-RNN) to better capture the temporal information contained in the future disturbance predictions.

\subsubsection{Reverse Time Recurrent Neural Networks}

Traditional recurrent neural networks (RNN) are a type of neural network that use a time-based structure to take advantage of temporal information in the data. RNNs take inputs from the current time step and from previous time steps that are passed through the RNN layer as a hidden state. RNNs perform significantly better than conventional feed-forward neural networks (FFNN) on sequential data applications such as forecasting and natural language processing. In our case, however, the input features do not contain data from previous time steps, but rather from future disturbances like weather, electricity price, and setpoint preferences. Nevertheless, future disturbances can also benefit from being used in RNNs, as is in the case of bidirectional RNNs, which use both previous and future datapoints to make a prediction at the current time step \cite{zeryer2017}. We apply this idea to Behavioral Cloning in the form of reverse-time RNNs, where the RNN is structured such that time is reversed, and future disturbance prediction information flows backward in time to help predict the optimal control at the current time step. 

\begin{figure}
    \centering
    \includegraphics[width=.7\textwidth]{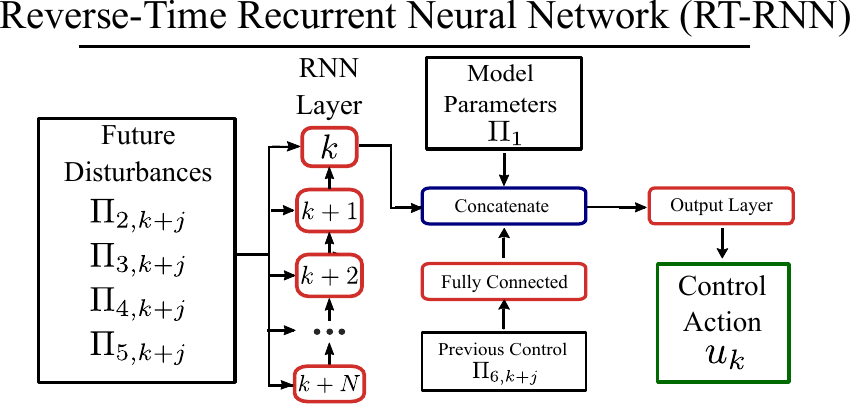}
    \caption{Reverse-Time Recurrent Neural Network Structure. Future disturbance parameter groupings containing weather, electricity price, and setpoint information are input into the reverse RNN layer where information flow backward in time over MPC horizon. These are concatenated with inputs from the other parameters and then to the output layer.}
    \label{fig: rt-rnn}
\end{figure}

Our proposed RT-RNN structure is given in Fig. \ref{fig: rt-rnn}. The parameter groupings that contain future disturbance information are input into the RNN layer. The information then flows backward in time, from the end of the MPC horizon to the current time step. The output of this layer is concatenated with the remaining input parameters. The previous control values indicating minimum on and off times contained in $\Pi_6$  are first passed through a single node layer to compress the information before concatenation. The final output layer contains a sigmoid activation function to give the binary control action prediction.

Since RNNs contain feedback loops to store memory, they can experience vanishing or exploding gradients if the sequences are too long. Thus, vanilla RNNs are often unable to learn long term temporal dependencies. To solve this, RNNs have been improved with model structures like gated recurrent units (GRU) and long short-term memory (LSTM), which are capable of storing a separate memory state that may be important in a long sequence. These structures can be equally applied for RT-RNNs, where the \textit{memory state} can instead be termed the \textit{prediction state}. For example, if a setpoint change occurs several hours in the future, the prediction state can store this information without it being potentially lost due to vanishing gradients over many time steps in the RNN. While LSTMs often outperform GRUs due to a more complex structure, GRUs can be more suited for memory constrained applications or on smaller datasets. Therefore, we test both model structures to determine their performance.

\subsection{Model Evaluation}

We evaluate the Behavioral Cloning agent in two steps: control prediction accuracy to determine the optimal Behavioral Cloning model structure and control simulation performance to determine its comparison to existing HVAC control policies.

\subsubsection{Model Structure Optimization}
\label{sec: grid_search}
We first evaluate control prediction accuracy to select the optimal model structure and hyperparameter configuration. We compare the control prediction performance and computational requirements of the RT-RNN to three more conventional supervised learning techniques: (1) FFNNs, (2) Random Forest, and (3) Extreme Gradient Boosting (XGBoost). FFNNs represent the most basic deep neural network architecture, and pass information forward from the input features to the output prediction through multiple fully connected layers. Each node in a layer contains a vector of weights for each of the nodes in the previous layer and a bias parameter. The value of each node is then put through a nonlinear activation function to allow the network approximate nonlinear functions.

Random forest is an ensemble based supervised learning method that uses an ensemble of many different decision trees to classify data. Different decision trees are fit based on random subsamples of the dataset, and each tree's output votes toward the final model's decision. By taking the majority vote of many decision trees, random forest reduces the potential for overfitting that is common with single decision trees. Both the memory requirement and performance of random forest depends on key hyperparameters that govern the number and size of the trees and must be optimized.

Extreme Gradient Boosting (XGBoost) is similar to random forest in that it uses an ensemble of decision trees, but it differs based on how the trees are created. Instead of creating each tree independently, XGBoost uses extreme gradient boosting to iteratively improve a decision tree using more trees. At each iteration, the algorithm constructs a new tree to predict the error resulting from the previous ensemble of trees and then adds the new tree to the ensemble using a scaling factor called the learning rate. By doing so, the algorithm "boosts" the prediction at each step until no more performance gains can be made. 

While more model parameters can theoretically learn more complex representations of the input data, this comes at the cost of larger model and higher memory requirements. To analyze this tradeoff, we determine each machine learning model's optimal hyperparameters through a grid search with 25 iterations for each model type. For each iteration, we log the model size and the validation prediction accuracy. Model size refers to the memory requirements to store each of the individual model parameters and is measured in kilobytes. Validation prediction accuracy refers to the model's prediction accuracy where the validation data is comprised of a random selection of 10\% of the buildings simulated in the training data.

\subsection{Control Simulation Performance}
After selecting the best predicting model, the actual control performance is found through control simulations. We define control performance as the cumulative MPC objective function over a five-day test simulation on a set of buildings $B$, operating conditions, and electricity tariffs that were not included in the original training dataset, represented by the equation, 
\begin{equation}
    \sum_{b=0}^{B} \sum_{k=0}^{K} \bigl[\pi_{e,k} P_k^b + \underline{\pi}_{\text{pen}}\underline{T}_{\text{pen},k}^b + \overline{\pi}_{\text{pen}}\overline{T}_{\text{pen},k}^b\bigr]
\end{equation}

Here $k$ is the time step and $K$ is the total number of time steps in the five-day test. Since setpoint preference and building thermal capacity can have a strong effect on MPC benefits, the model is tested on ten different buildings indexed by $b$ to give a more holistic evaluation of model performance and generalization. Final computational requirements are logged during this simulation and include the processing speed and memory requirements required to store and run the model.

We use these metrics to compare the Behavioral Cloning control to a baseline standard rule-based control policy and the true MPC policy. In this case, the rule-based control policy is the typical thermostat's hysteresis control, where the heat pump turns on when the indoor temperature falls below the lower comfort bound and turns off when the temperature rises above the upper comfort bound. Note that this rule-based policy uses variable setpoint schedules that may include energy-saving setbacks when the occupant is away or asleep. In contrast, the MPC policy provides the target objective function value that Behavioral Cloning is trying to imitate. 

\section{Model Training Results}
\label{sec: results}

\subsection{DAgger Training Data Generation}

At each iteration of the DAgger algorithm, the system simulates new buildings with different random $ R $, $ C $, and $ \alpha $ values and different heat pump performance coefficients. Various setpoint schedules were obtained from the Ecobee Donate Your Data dataset \cite{ecobee}, which contains smart thermostat setpoint schedules from thousands of homes throughout the country. Thermal comfort band schedules were set based on whether those thermostats were in "home", "sleep", or "away" modes. We assume the comfort band is $\pm$ $.5 ^\circ$C for "home", $\pm$ $1.0 ^\circ$C for "sleep", and no limit when "away". Electricity price schedules were obtained from New York State Electric and Gas (NYSEG) \cite{NYSEG}, ConEdison (ConEd) \cite{CONED}, and Xcel Energy \cite{xcel}, three utilities that offer time-of-use rates during winter. Weather data comes from various days in January and February 2019 for New York City \cite{NSRDB}. Note that while our method provides some level of generalization, if the climate varies significantly from training case, more simulations specific to the target climate may be required.

We generated 15 days of simulation data, each containing 10 randomized buildings, heat pumps, and setpoint schedules. For each of the random buildings, the thermodynamic model parameters were randomly selected from a range of $\pm 25\%$ around the values used in \cite{Lee2020a}. This totals to 45,760 samples of data used for training. To show the benefit of both the constraint-informed parameter normalization and the DAgger algorithm, we trained a set of models on three different training data representations. The first (CIPG + DAgger) is the aforementioned dataset generated by DAgger and normalized using our constraint-informed parameter groupings. The second (CIPG + AMPC) uses our constraint-informed parameter groupings, but instead is trained to approximate the closed loop MPC simulations and thus contains no information outside of the optimal control trajectory. The third (No Parameter Groupings + DAgger) contains data from the DAgger-generated dataset that is independently normalized. In other words, the third dataset uses only the conventional machine learning approach of scaling each individual input variable to have zero mean and unit variance, rather than our approach of first creating CIPGs and then scaling. 

\begin{figure}
    \centering
    \includegraphics[width=.7\textwidth]{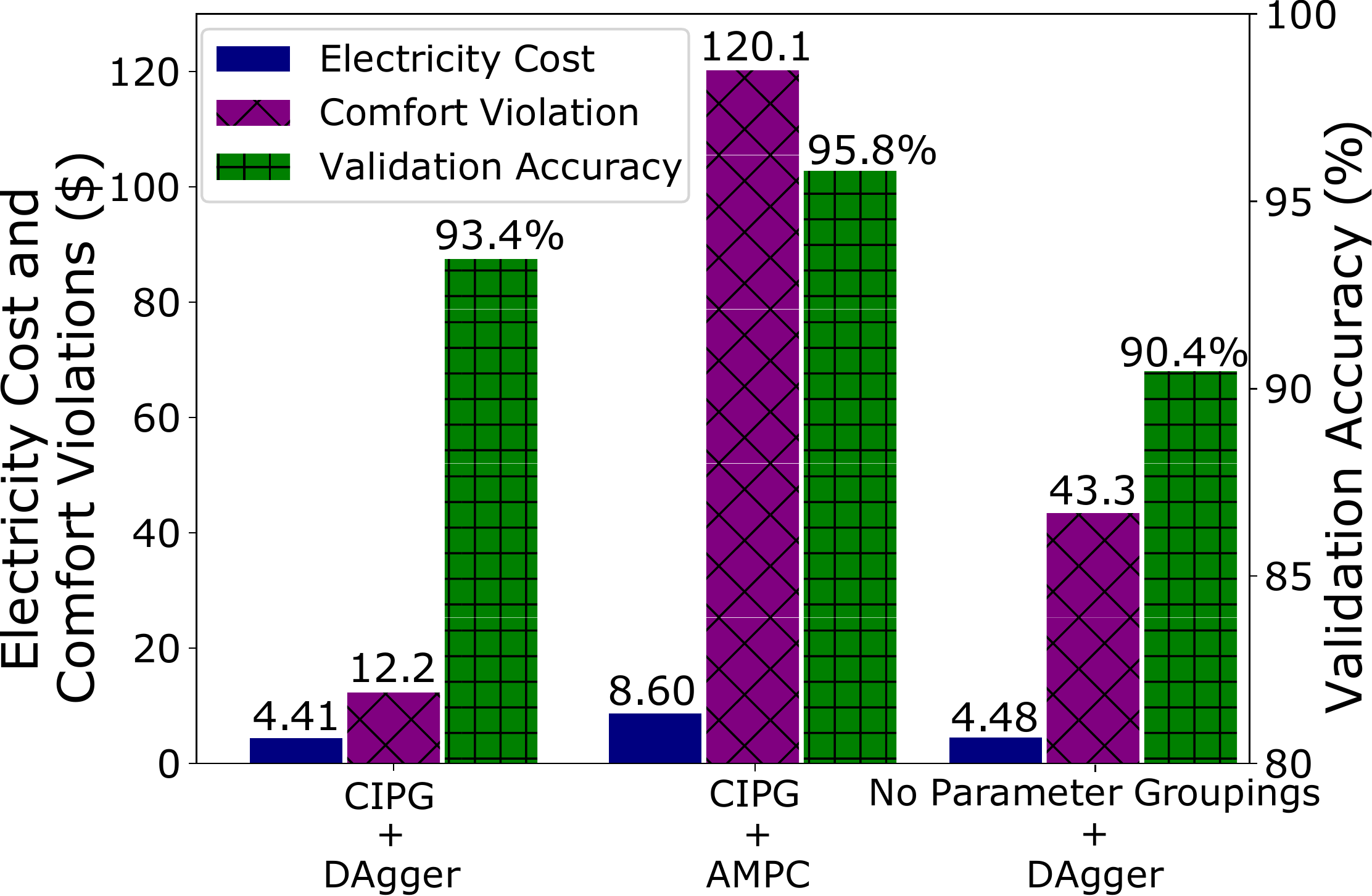}
    \caption{Per-unit Control Objectives (lower is better) and Validation Prediction Accuracy (higher is better) for (1) constraint informed parameter groupings (CIPG) with DAgger training data, (2) CIPG and trained to approximate closed loop MPC (AMPC), and (3) CIPG with DAgger training data. When combined, our contributions, CIPG and DAgger, provide more stability and lower costs.}
    \label{fig: dataset_effect}
\end{figure}

To compare the datasets, we trained 25 RT-RNNs for each dataset using various hyperparameter combinations to find the combination that provided the highest prediction accuracy on validation data. We then tested each dataset's best model in a control simulation containing new conditions outside of the training dataset. Control performances for each dataset are shown in Fig. \ref{fig: dataset_effect}. 

There are two important findings from these results. First, combining the features into parameter groupings in CIPG + DAgger provides a three percentage point increase in validation prediction accuracy over No Parameter Groupings + Dagger, meaning it improves the ability to fit the dataset without overfitting. While there is no significant difference in electricity cost, the improved prediction accuracy translates to significantly reduced comfort violations. Second, despite lower validation accuracy, Behavioral Cloning trained with DAgger has an order of magnitude better control performance than the AMPC model, which was trained to approximate closed loop MPC. The higher accuracy on the AMPC dataset is somewhat misleading and does not translate to better control performance. Since it is trained on closed loop MPC simulations the data is more homogeneous, and the indoor temperature is always within the thermal comfort limits. This contrasts with the DAgger dataset, which has data across a range of indoor temperatures, particularly from early iterations when the model does not perform well. The implication is that while it is easier to fit a more homogenous dataset, the AMPC model has insufficient data to correct itself if it strays from the optimal trajectory, and the result is a model with no knowledge that comfort violations are undesirable. 

\subsection{Optimal Behavioral Cloning model structure}

\begin{figure}[h]
    \centering
    \includegraphics[width=.65\textwidth]{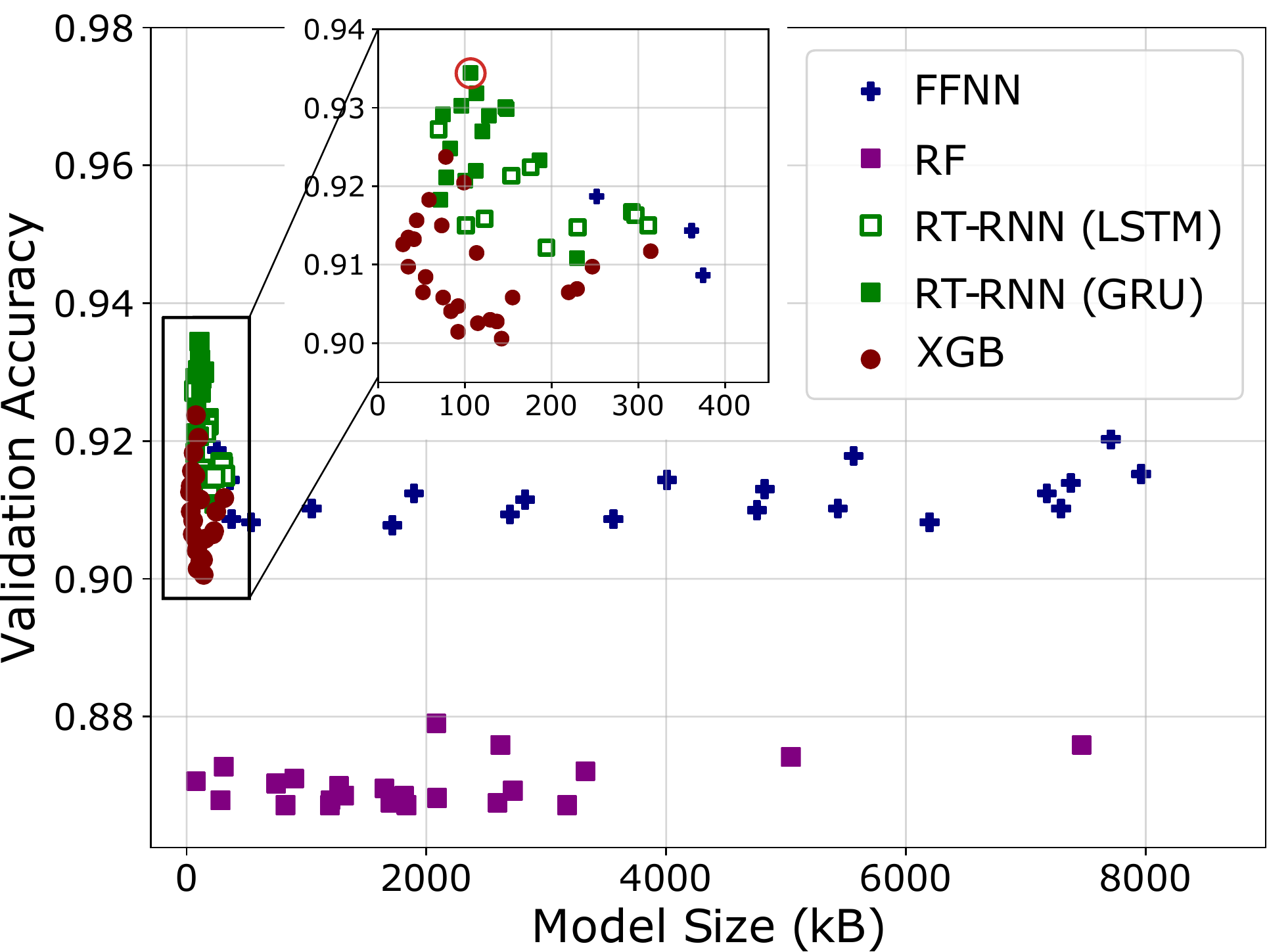}
    \caption{Validation accuracy versus model size for each of the four tested machine learning algorithms. The reverse-time RNN (RT-RNN) models largely outperform the other models on both metrics: It can maintain the highest prediction accuracy with a very small memory requirement. The circled marker denotes the chosen model.}
    \label{fig: model selection}
\end{figure}

Fig. \ref{fig: model selection} gives the results of the hyperparameter grid search in terms of validation accuracy and model size as presented in Sec. \ref{sec: grid_search}. The worst performers were the feed-forward neural network and random forest, each requiring high memory requirements with only marginal performance increases from more complex models. XGBoost and the LSTM Reverse-time Recurrent Neural Network (RT-RNN) performed similarly, while the GRU RT-RNN performed the best. Therefore, for our final Behavioral Cloning agent we chose the GRU RT-RNN configuration with the highest validation prediction accuracy encircled in Fig. \ref{fig: model selection}.

The optimal model configuration for the selected RT-RNN encircled in Fig. \ref{fig: model selection} contains one GRU layer with 26 nodes and 6 channels corresponding to each of the parameter groupings that are indexed over the MPC control horizon ($\Pi_2$ through $\Pi_7$). The previous control values ($\Pi_5$) are input to the model through a 1-node layer with ReLu activation function \cite{glorot2011deep}. The outputs of these layers are concatenated with the building model parameters ($\Pi_1$) and connected to a 25 node fully connected layer with ReLu activation function. It is then connected to the output layer with sigmoid activation to give the binary control value prediction. Other training hyperparameters are summarized in Tab. \ref{tab: opt network}

\begin{table}[h]
\caption{Results and training parameters for the optimal RT-RNN configuration}
\label{tab: opt network}
\begin{center}
\small
\begin{tabular}{p{0.3\textwidth}p{0.1\textwidth}}
\toprule
Model Type & GRU \\
\midrule
Batch Size & 512 \\
Optimizer & Adam \\
Training Epochs & 24 \\ 
Model Size & 106 kB\\
Validation Accuracy & 94.5\% \\

\bottomrule
\label{tab: opt network}
\end{tabular}
\end{center}
\end{table}

\subsection{Final Control Model Results}

Using the selected optimal RT-RNN model configuration, we analyzed the control performance compared to a baseline thermostat control and the target true MPC control. Tab. \ref{tab: final results} gives the average processing time, memory requirements, and the average per building electricity cost and comfort violation on the test conditions. The simulations were computed on a Raspberry Pi Zero, which contains a 1GHz single core processor with 512 MB of RAM. Behavioral cloning only requires .1\% of the memory of MPC and can operate around 93,000x faster, all while maintaining a similarly low electricity cost and only a modest increase in comfort violations. Moreover, on average the Raspberry Pi, which contains more computing hardware than a typical PLC, was unable to even solve the MPC within the required time step (300 seconds).

\begin{table}[h]
\caption{Control Performance for 10 buildings over a five-day span in February computed on a Raspberry Pi Zero.}
\label{tab: params}
\begin{center}
\small
\begin{tabular}{p{0.15\textwidth}p{0.17\textwidth}p{0.1\textwidth}p{0.1\textwidth}p{0.17\textwidth}p{0.16\textwidth}}
\toprule
Model & Objective Value & Electricity Cost & Comfort Violation & Memory Requirement & Computational Time Per Step \\
\midrule
Rule-Based (baseline)    & 270.79  & \$49.45  & 221.34 & $\sim$ 0          & 5e-4 s \vspace{1mm}\\
Behavioral Cloning                   &  161.12 & \$42.31  & 118.81 & 176 kB     & 3.3e-3 s \vspace{1mm}\\
MPC (target)             &  144.04 & \$ 42.46   & 101.58  & 150,000 kB & 309 s \\

\bottomrule
\label{tab: final results}
\end{tabular}
\end{center}
\end{table}

\begin{figure}[h!]
    \centering
    \begin{subfigure}[b]{0.49\textwidth}
        \centering
        \includegraphics[width=\textwidth]{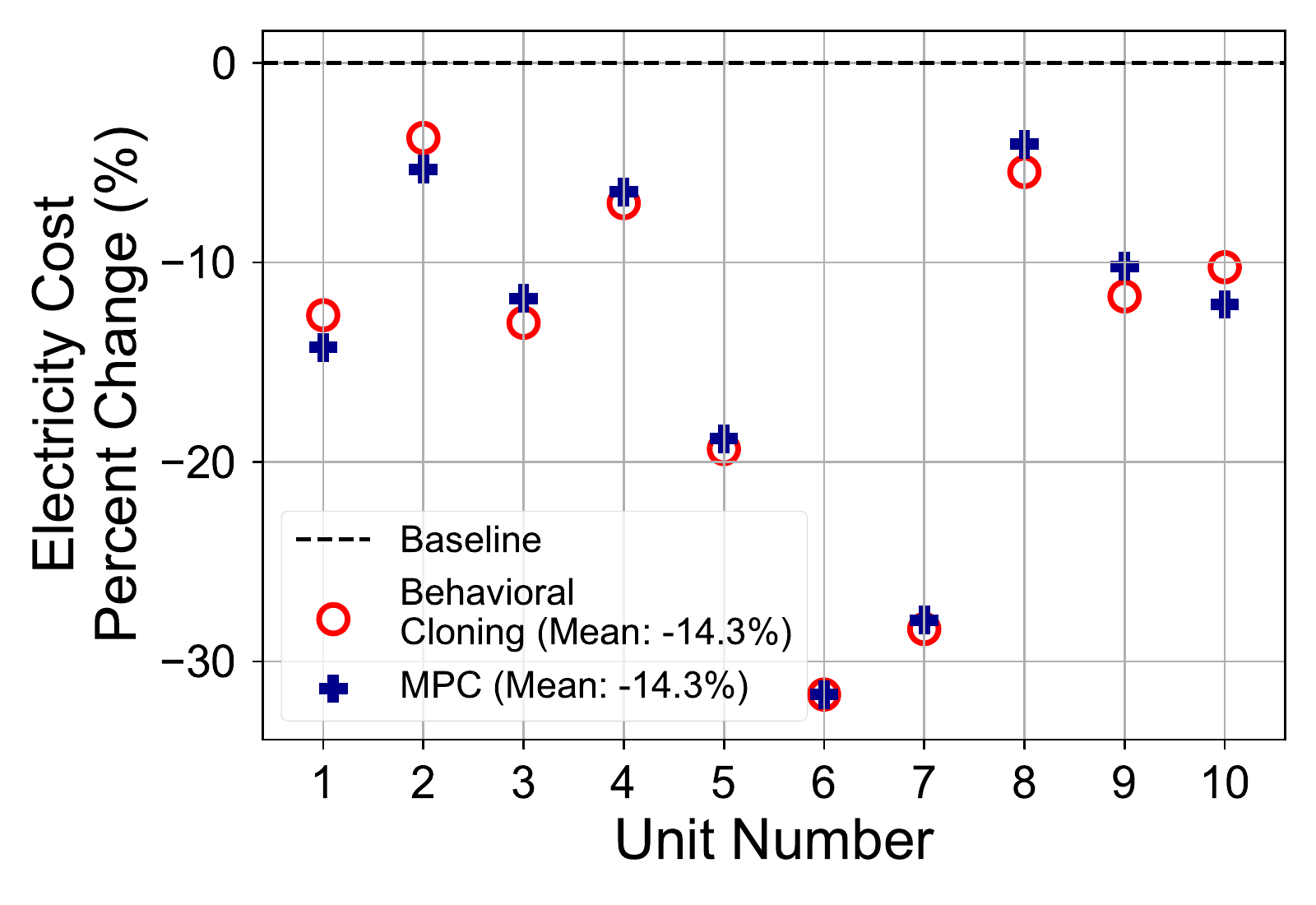}
        \caption{}    
    \end{subfigure}
    \hfill
    \begin{subfigure}[b]{0.49\textwidth}  
        \centering 
        \includegraphics[width=\textwidth]{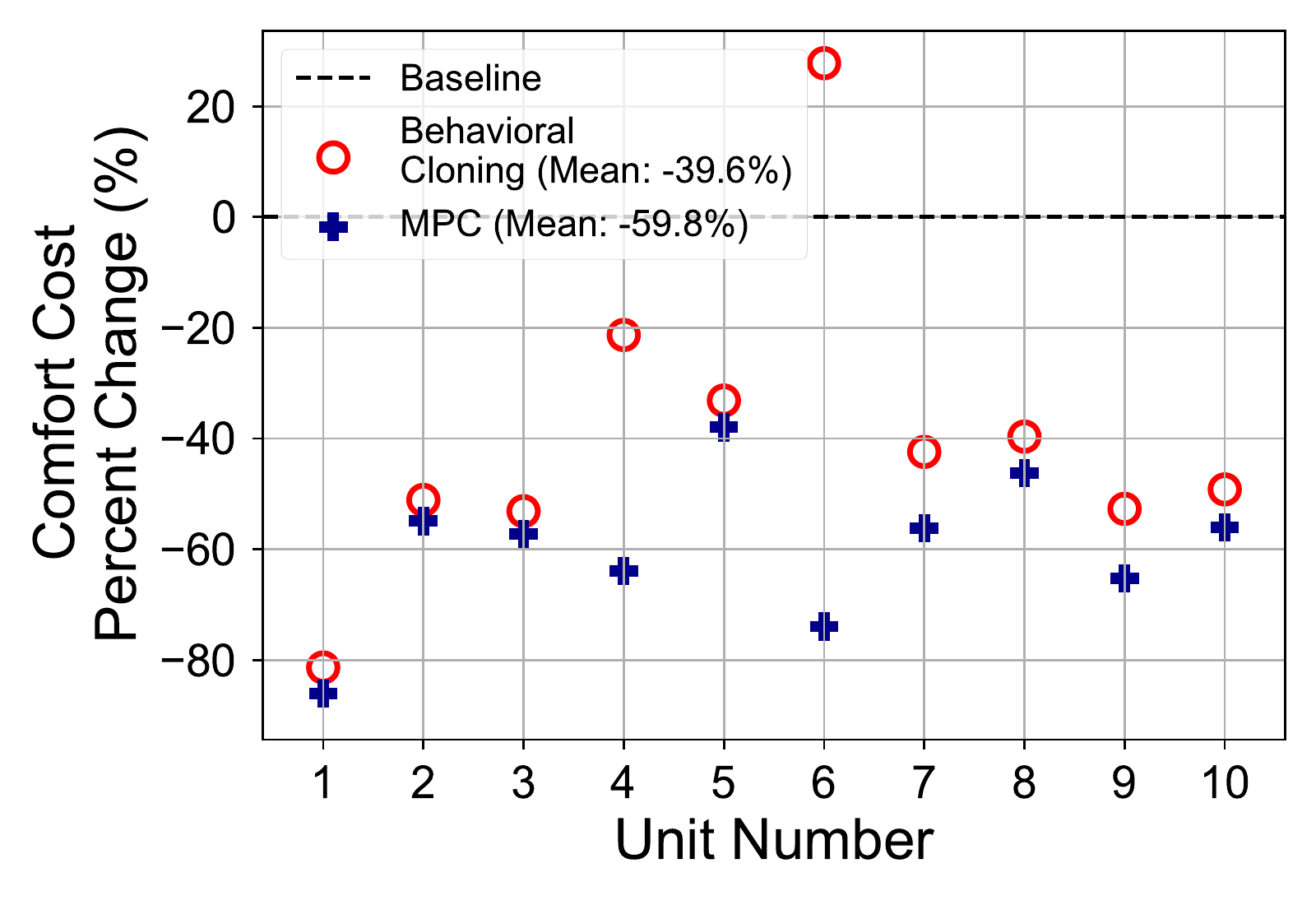}
        \caption{}       
    \end{subfigure}
    \caption{Despite requiring orders of magnitude less computational time, objective values improvements over the baseline rule-based approach for each building using Behavioral Cloning show similar performance improvements to that of MPC. Differences in the benefits between buildings are due to varying setpoint schedules and the level of building insulation.}
    \label{fig: per unit}
\end{figure}

Fig. \ref{fig: per unit} depicts each building's percent improvement in electricity cost and thermal comfort for Behavioral Cloning and MPC compared to the baseline rule-based approach. On average, MPC and Behavioral Cloning perform similarly, with broad improvements to both electricity cost and thermal comfort compared to the baseline. These improvements can vary significantly from building to building based on the setpoint schedules and how well the building is insulated. Note that the large percent increase in comfort violation for Building 5 is due to a very small baseline comfort violation, and the magnitude of increase is under $0.25$.

\begin{figure}
    \centering
    \includegraphics[width=\textwidth]{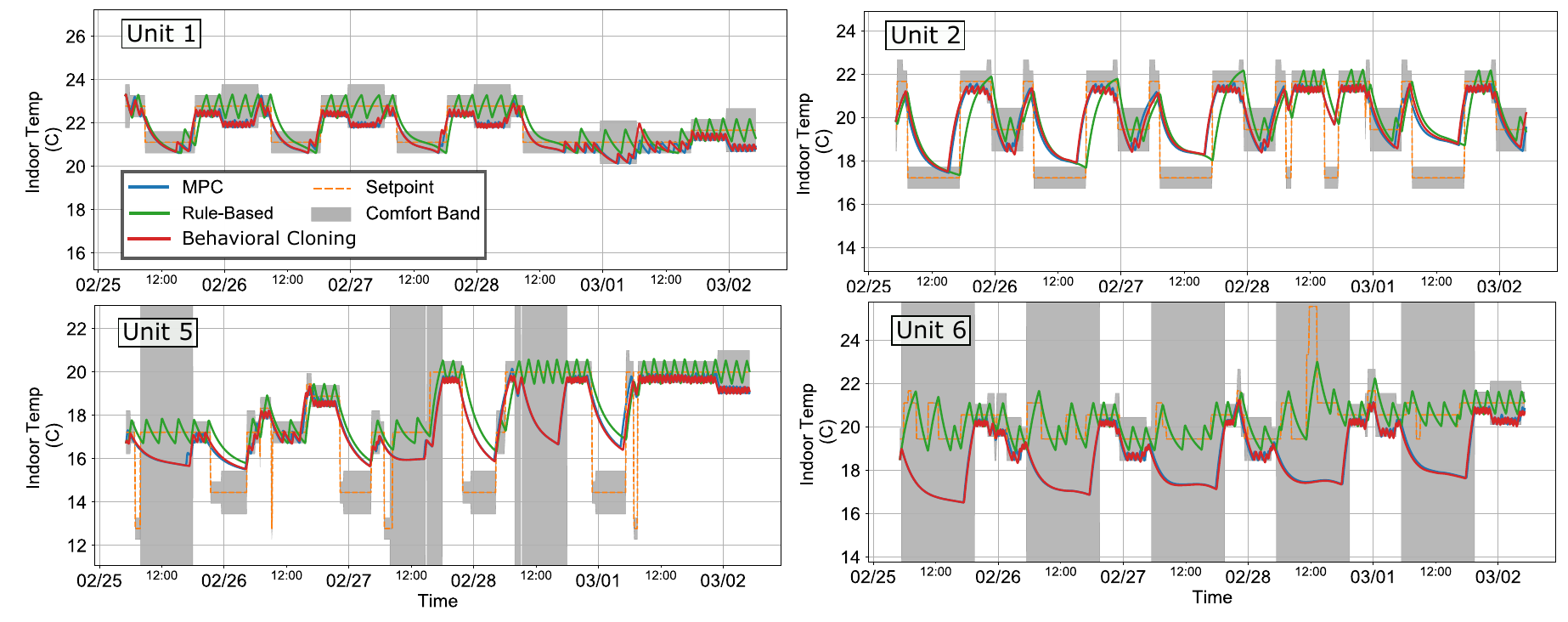}
    \caption{Test control plots for a representative sample of the buildings under each control method. Both MPC and Behavioral Cloning can more effectively take advantage of varying setpoint schedules and comfort bands by reducing consumption during times the resident is away, and optimally preheating to avoid comfort violations.}
    \label{fig: control results}
\end{figure}

Fig. \ref{fig: control results} presents the temperature trajectories for a representative sample of the buildings for each of the control policies: baseline rule-based control, Behavioral Cloning, and MPC. This sample shows the various operating conditions that occur in the overall simulation: small and large setpoint changes and small and large amounts of time when the resident is away. Similar to MPC, Behavioral Cloning maintains the temperature within the lower range of the acceptable thermal comfort band, while still able to effectively preheat the building in preparation for large setpoint changes. These control plots emphasize that though the Behavioral Cloning does not contain any explicit thermodynamic equations or solve any optimization problem, it is able to generalize to new operating conditions and changing user preferences like that of MPC. Each of these setpoint schedules and building-heat pump thermodynamics were not originally included in the training dataset.

\section{Conclusion}
\label{sec: conclusion}
In this paper, we have presented a highly scalable and easy-to-install method for implementing Behavioral Cloning of model predictive control (MPC) on low-cost hardware in many different residential buildings. Our method significantly reduces the installation effort and cost compared to previous approximate MPC studies by allowing the Behavioral Cloning agent to be trained once for many buildings and operating conditions, rather than needing to be retrained for each specific building it will be implemented on. In addition, our method can adapt to new setpoint schedules and different time-of-use electricity prices, which consistently occur in online operation. 

Simulation results across a range of building parameters, setpoint schedules, and electricity price schedules show that our method provides identical average efficiency improvements to that by MPC, with a small increase in comfort violations. However, the comfort violations of Behavioral Cloning are still far below that of the baseline rule-based approach. Finally, our method only requires .1\% of the memory requirements of conventional MPC and can provide the optimal control around 93,000x faster, drastically reducing the computational hardware cost for implementation.

Encouraging building owners to retrofit fossil-fuel systems in favor of heat pumps and to adopt smart building climate control has been, and will likely continue to be, a challenging problem. High capital costs combined with building owners' lack of sufficient knowledge act as barriers to more widespread adoption of clean and efficient heating and cooling. Our method for Behavioral Cloning of MPC can potentially mitigate these barriers by providing a low-cost plug-and-play solution for efficient and flexible heating and cooling control.

\section*{Acknowledgments}
The authors acknowledge the support from the National Science Foundation (NSF) under grant 1711546, the NSF Graduate Research Fellowships Program (to ZEL) and the Cornell Atkinson Center for Sustainability.

\bibliography{mybibfile}
\end{document}